\documentclass[12pt]{article}
\usepackage{epsf,afterpage}
\usepackage{amsfonts}
\newcommand{\be}{\begin{equation}}
\newcommand{\ee}{\end{equation}}
\def\la{\mathrel{\mathpalette\fun <}}
\def\ga{\mathrel{\mathpalette\fun >}}
\def\fun#1#2{\lower3.6pt\vbox{\baselineskip0pt\lineskip.9pt
\ialign{$\mathsurround=0pt#1\hfil
##\hfil$\crcr#2\crcr\sim\crcr}}}
\newcommand{\ven}{\mbox{\boldmath${\rm n}$}}
\newcommand{\bc}{\begin{center}}
\newcommand{\ec}{\end{center}}

\newcommand{\mr}[1]{\mathrm{#1}}
\newcommand{\lt}{\left}
\newcommand{\rt}{\right}
\newcommand{\lb}{\label}
\newcommand{\fr}[2]{\frac{#1}{#2}}
\newcommand{\lan}{\left\langle}
\newcommand{\ran}{\right\rangle}
\newcommand{\tr}{\mathrm{tr}}

\title{\bf Static potential in baryon in the method of  field
correlators}
\author{D.S.Kuzmenko}

\date{\it State Research Center\\  Institute of Theoretical and
Experimental Physics, \\
 117218, B. Cheremushkinskaya 25, Moscow, Russia}

\begin{document}
\maketitle
\begin{abstract}
The static three-quark potential in arbitrary configuration of
quarks is calculated analytically. It is shown to be in a full
agreement with the precise  numerical simulations in lattice QCD.  The
results of the work have important application in nuclear physics, as
they allow to perform accurate analytic calculations of spectra
of the baryons.
\end{abstract}

\section{Introduction}

Precise knowledge of the nucleon spectroscopy is very important
for the deep understanding of the nuclear processes. The
nonperturbative static potential is a key quantity in calculations
of the baryon spectra, and this is why its investigation is one of the
important problems in nuclear physics.
Besides, the study of the static potential provides an insight into
the two fundamental properties of strong interactions ---
confinement of the color charges and  scale of the fluctuations
of the confining gluonic fields.

In the present paper static potential in baryon is calculated
using the method of the field correlators (MFC) \cite{correlators},
which is based directly on QCD and operates with the vacuum averages
of the correlators of the gluonic field strengths.

It is suggested in MFC that the gluonic fields play a twofold role in
QCD. First, gluons propagate dynamically in the vacuum, and this
process at small distances can be described by the pertubation theory.
In particular, the interaction of the quarks due to
the one-gluon-exchange leads to the color-coulomb interaction
potential.
Second, gluons in the vacuum form the nonperturbative condensate,
which constitutes the background, where the perturbative gluons
propagate.

Two-point (or bilocal) correlators of gluonic fields are parameterized
in MFC by the scalar formfactors \cite{MFC}. This parameterization
gives rise to the area law for the Wilson loop at sufficiently large
separations between color sources, which ensures confinement. The
slope $\sigma$ of the linear static potential between quark and
antiquark, corresponding to the Wilson loop, is the main parameter of
MFC. Using this the only parameter one can obtain in MFC the spectra
of the mesons within the 5\% accuracy \cite{Regge}. The value of
$\sigma$ is determined phenomenologically from the slope of the
Regge trajectories of mesons, and corresponds to the confinement
radius.

The scalar formfactors of the background fields fall off
exponentially, which reflects the stochastic nature of the confining
fluctuations of the gluonic background field. The correlation length
of the background fields $T_g$ is sometimes considered as the second
parameter of the MFC, although it can generally be expressed through
the $\sigma$ \cite{glump}. Its value is small in comparison with the
radius of confinement and negligible for mesons. The situation for
baryons is different because of the correlations of the background
fields near the string junction, which lead according to MFC to the
decreasing slope of the potential at small and intermediate quark
separations. There is another $T_g$-induced effect, which is
considered in the present work, namely the potential difference
between the configurations with the same length of the string, but
different locations of quarks.

The  appearence of recent accurate numerical calculations of the
static three-quark potential in lattice QCD in quenched approximation
\cite{Takahashi,deForcrand} allow to test analytic results of MFC.
We perform this important test in the paper and find a complete
agreement between MFC and lattice results.

Before to calculate the static potential, in the next section of the
paper we consider the Wilson loop of the baryon, which in the case of
the static quarks is the baryon Green function.
We express vacuum average of the Wilson loop through the  vacuum
averages of bilocal correlators of the gluon field strength tensor
and parameterize them according to MFC. A brief discussion of the
 gluon condensate value obtained from the ITEP sum rules and its
relation with the MFC is also given.
In third section the procedure of calculation of the potential is
described, and  its behavior is analized
in comparison with the lattice data. In the concluding section a
summary of the results obtained in the paper is given, and their
perspective applications are discussed.

\section{The baryon Wilson loop and its average in MFC}

It is known that the gauge-invariant state of the
baryon with quarks at points $x,~y,~z$ reads as
\be
\Psi_B(x,y,z,Y) =
\epsilon_{\alpha\beta\gamma}\Phi^{\alpha}_{\alpha '}(x,Y,C_1)
\Phi^{\beta}_{\beta '}(y,Y,C_2)\Phi^{\gamma}_{\gamma '}(z,Y,C_3)
q^{\alpha '}(x) q^{\beta '}(y)q^{\gamma '}(z),
\lb{B1}
\ee
where $\alpha, ~\beta,...$ are the color indexes of the fundamental
representation of  $SU(3)$,
\be
\Phi^{\alpha}_{\beta}(x,y,C)= (P\exp ig\int_C A_\mu d
z_\mu)^{\alpha}_{\beta}
\lb{B2}
\ee
is a parallel transporter along the contour $C$, which  connects
points $x,~y$ ($P$ in (\ref{B2}) means the ordering of color
matrices along the trajectory of integration); $q$ and $A$
are quark and gluon field operators, and $Y$ is the point of the
string junction. The baryon Green function is defined as a vacuum
average
\be
 G_B(\bar X, X)=\lan \Psi_B^+ (\bar X)\Psi_B(X)\ran,
\label{B3}
\ee
where $X\equiv x,y,z,Y$.
If quarks are static, the Green function becomes the Wilson
loop ${\cal W}_B$, which reads as
\be
{\cal W}_B=\lan \frac16 \epsilon_{\alpha\beta\gamma}
\epsilon^{\alpha '\beta '\gamma '}
\Phi^{\alpha}_{\alpha '}(C_1)\Phi^{\beta}_{\beta '}(C_2)
\Phi^{\gamma}_{\gamma '}(C_3)\ran.
\label{B4}
\ee

\begin{figure}[!t]
\epsfxsize=8cm
\hspace*{4.35cm}
\epsfbox{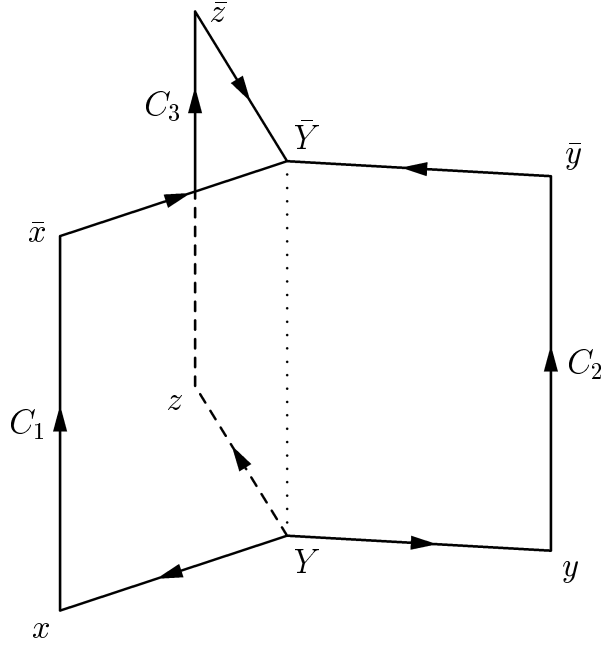}
\caption{Baryon Wilson loop}
\label{YWloop}
\end{figure}

The contours $C_1,~C_2,~C_3$ and direction of the integration are
shown in Fig. 1. The trajectory of the string junction, which we
denote  $C_0$, is shown in the Fig. 1 by the dotted line.
 Note that the definition (\ref{B4}) of the baryon
Wilson loop is generally accepted. The static potential in baryon is
related with the Wilson loop as follows,
\be
V_B=
-\lim_{T\to \infty} \frac{1}{T}\ln  {\cal W}_B,
\label{B5}
\ee
where $T$ is the time extension of the Wilson loop.

We are now to express the Wilson loop (\ref{B4}) through the
correlators of the gluon field strength tensor. For this sake
rewrite it as
\be
{\cal W}_B=\lan \frac16 \epsilon_{\alpha\beta\gamma}
\epsilon^{\alpha '\beta '\gamma '}
\Phi^{\alpha}_{\alpha '}(\tilde C_1)\Phi^{\beta}_{\beta '}(\tilde C_2)
\Phi^{\gamma}_{\gamma '}(\tilde C_3)\ran,
\label{B6}
\ee
where $\tilde C_i=C_i\cup C_0$. The expression (\ref{B6})
follows from (\ref{B4}) after the substitution the latter with the
equation
\be
\epsilon^{\alpha '\beta '\gamma '}=
\epsilon^{\alpha ''\beta ''\gamma ''}
\Phi^{\alpha''}_{\alpha '}(C_0)\Phi^{\beta''}_{\beta '}(
C_0) \Phi^{\gamma''}_{\gamma '}(C_0).
\label{B7}
\ee
Note that the last relation is valid in the generalized coordinate
gauge, in which $\Phi^{\alpha}_{\alpha'}(C_0)=
\delta^{\alpha}_{\alpha'}$, but the expression (\ref{B6}) is a scalar
of the color $SU(3)$ and therefore does not depend on the choice of
the gauge. Since the integration in (\ref{B6}) is performed along
three closed contours, one may use for these contours of the baryon
Wilson loop the nonabelian Stokes theorem \cite{correlators}. As a
result one gets
\be
{\cal W}_B=\lan \frac16 \epsilon_{\alpha\beta\gamma}
\epsilon^{\alpha '\beta '\gamma '}
W^{\alpha}_{\alpha '}(S_1)W^{\beta}_{\beta '}(S_2)
W^{\gamma}_{\gamma '}(S_3)\ran,
\label{B8}
\ee
where $S_i$ are the minimal surfaces of the contours $\tilde C_i$,
and
\be
W^{\alpha}_{\alpha '}(S)=
{\cal P}\exp(ig\int_S
d\sigma_{\mu\nu}(x)F_{\mu\nu}(x)\Phi(x,x_0))^{\alpha}_{\alpha'}
\label{B9}
\ee
does not depend on the choice of the point $x_0$, if the latter places
at the surface $S$ \cite{correlators}. Expand now the exponent in
(\ref{B9}),
\be
W^{\alpha}_{\alpha '}(S)=
\delta^{\alpha}_{\alpha'}+
ig\int_S d\sigma\, F^{\alpha}_{\alpha'}-
\frac{g^2}2\int_S\int_S d\sigma(1)\, d\sigma(2)\,
F^{\alpha}_{\beta}(1)  F^{\beta}_{\alpha'}(2)+...,
\label{B10}
\ee
where we have explicitely written only the color indexes and denoted
by the dots the higher powers of the expansion.
Since the vacuum average of the strength tensor equals to zero, the
Wilson loop will read as follows,
$$
{\cal W}_B=\frac16 \epsilon_{\alpha\beta\gamma}
\epsilon^{\alpha '\beta '\gamma '}\left(
\delta^{\alpha}_{\alpha'}\delta^{\beta}_{\beta'}
\delta^{\gamma}_{\gamma'}-
\delta^{\beta}_{\beta'}\delta^{\gamma}_{\gamma'}\,
\frac{g^2}2\int_{S_1}\int_{S_1} d\sigma(1)\, d\sigma(2)\,
F^{\alpha}_{\rho}(1)  F^{\rho}_{\alpha'}(2)-\right.
$$
\be
-\left.\delta^{\gamma}_{\gamma'}\,
g^2\int_{S_1}\int_{S_2} d\sigma(1)\, d\sigma(2)\,
F^{\alpha}_{\alpha'}(1) F^{\beta}_{\beta '}(2)+...\right),
\label{B11}
\ee
where the dots denote the double integrals over surfaces
 $S_2S_2,~S_3S_3,~S_1S_3$, and $S_2S_3$, analogous to the written
ones, as well as the higher order correlators. Taking into account
that $\epsilon_{\alpha\beta\gamma}\epsilon^{\alpha '\beta\gamma}
F^{\alpha}_{\rho} F^{\rho}_{\alpha'}=2\,\tr\, FF$,
$\epsilon_{\alpha\beta\gamma}\epsilon^{\alpha '\beta '\gamma}
F^{\alpha}_{\alpha'} F^{\beta}_{\beta '}=-\tr\, FF$, one gets from
(\ref{B11})
$$
{\cal W}_B=\exp\left\{-\sum_{i=1}^3\frac12\int_{S_i}\int_{S_i}
 d\sigma(1) d\sigma(2)\lan\frac{g^2}3\tr F(1)F(2)\ran+\right.
$$
\be
\left.+\sum_{i<j}\frac12\int_{S_i}\int_{S_j}
 d\sigma(1) d\sigma(2)\lan\frac{g^2}3\tr F(1)F(2)\ran\right\}.
\label{B12}
\ee
Here and in what follows we disregard the higher order correlators,
assuming that their contribution to Wilson loop is small or
can be reduced to the contribution of the bilocal ones. This
assumption leads to the area law for the Wilson loop and is
shown in \cite{Gauss} to be motivated by the  property of the Casimir
scaling of the potentials of static sources in  various
representations of $SU(3)$.

The second motivation of the cancellation of the contributions
of higher order correlators to the Wilson loop follows from  the
stochastic properties of the QCD vacuum. Note that the bilocal
correlators written in (\ref{B12}) are the first term of the cluster
expansion used in the theory of fluctuations \cite{cluster} for the
description of the correlated stochastic processes. According to the
fundamental identity of the cluster expansion, higher order
correlators enter into (\ref{B12}) as cumulants, or connected
correlators (one can find an accurate definition of cumulants as well
as the proof of the fundamental identity in \cite{cluster}), which
rapidly fall off with the increasing order for almost independent
points. That's why the bilocal approximation taking into
account only two-point correlators is motivated by the theory of
fluctuations at distances greater than the correlation length of the
gluon fields.

The bilocal correlators are parameterized in MFC  \cite{MFC} in the
case of $SU(N_c)$ using two scalar formfactors $D$ and $D_1$ as
follows,
$$
    \frac{g^2}{N_c}\tr \langle F_{\mu_1\nu_1}(x)\Phi(x,x')
F_{\mu_2\nu_2}(x')   \Phi(x',x)\rangle=
(\delta_{\mu_1\mu_2}\delta_{\nu_1\nu_2}-
\delta_{\mu_1\nu_2}\delta_{\mu_2\nu_1})D(z)+
$$
\be
   +\frac12\lt(\fr{\partial}{\partial z_{\mu_1}}
(z_{\mu_2}\delta_{\nu_1\nu_2}-z_{\nu_2}\delta_{\nu_1\mu_2})+
\fr{\partial}{\partial z_{\nu_1}}
(z_{\nu_2}\delta_{\mu_1\mu_2}-z_{\mu_2}\delta_{\mu_1\nu_2})\rt)D_1(z)
\equiv{\cal D}_{\mu_1\nu_1,\mu_2\nu_2}(z),
\label{B13}
\ee
where $z\equiv x-x'$. The formfactors of the background fields read as
\be
D(z) = D(0) \exp \lt(-\fr{|z|}{T_g}\rt),
~~D_1(z) = D_1(0) \exp \lt(-\fr{|z|}{T_g}\rt).
\label{B14}
\ee
The exponential behavior of the formfactors is confirmed by the
lattice simulations \cite{latDD1,BBV} at distances $z\ga 0.2$ fm.
The values of the parameters obtained are $T_g= 0.12\div 0.2$ fm
\cite{latDD1,BBV} and  $D_1(0)/D(0)\approx 1/3$ \cite{latDD1}.

In the case of static quark and antiquark the relation analogous to
 (\ref{B12}) reads as
\be
{\cal W}_B=\exp\left\{-\frac12\int_{S}\int_{S}
 d\sigma(1) d\sigma(2)\lan\frac{g^2}{N_c}\tr F(1)F(2)\ran\right\}
\label{B15}
\ee
and at large quark separations $R\gg T_g$ gives rise to the area law
for the Wilson loop,
\be
  \langle {\cal W} \rangle^{\mathrm{biloc}}=
\exp(-\sigma S),
\label{B16}
\ee
where the string tension $\sigma$ is given by the relation
\be
\sigma=\fr{\pi}2\int_0^\infty dz^2\,D(z).
\label{B17}
\ee
The value of the string tension $\sigma\approx$ 0.18 GeV$^2$
may be determined phenomenologically from the slope of the meson Regge
trajectories \cite{Regge}. Note that the value $T_g\approx 0.13$ fm
can be extracted from the gluelump spectrum, which is calculated in
MFC \cite{glump} using the only nonperturbative parameter $\sigma$.

From (\ref{B13}) and (\ref{B14}) the equation
\be
\lan\frac{\alpha_s}{\pi}F_{\mu\nu}^a(x) F_{\mu\nu}^a(0)\ran=
\frac{18 D(x)}{\pi^2}
\label{B18}
\ee
 follows, which relates in particular the gluon condensate
$\lan\frac{\alpha_s}{\pi}F_{\mu\nu}^a(0)F_{\mu\nu}^a(0)\ran
\equiv\lan\frac{\alpha_s}{\pi}F^2\ran$ with the value of the
formfactor  $D$ at zero.
Although $D(0)$ does not enter formally into the definition of
$\sigma$ (\ref{B17}), but assuming the behaviour of $D(z)$ (\ref{B14})
at small distances (this assumption is rigorously speaking false
\cite{zero}, nevertheless we use it in this paper as an
approximation; the accurate expressions at small distances can be
found in \cite{baryon}), one gets $\sigma=\pi D(0)T_g^2$, which allows
to evaluate the gluon condensate as follows,
 $\lan\frac{\alpha_s}{\pi}F^2\ran=
18\,\sigma/(\pi^3 T_g^2)\approx 0.25$ GeV$^4$, for $T_g=0.13$ fm.

The value of the gluon condensate can also be obtained from the ITEP
sum rules (see the book \cite{sumrules} and references to the original
works therein). The sum rules use the operator expansion for the
currents, and the gluon condensate contribute to it in combination
$\lan\frac{\alpha_s}{\pi}G^2\ran/Q^4$. Herewith the value of the gluon
condensate is assumed to be independent on the momentum.
Recently on the basis of the analysis using the sum rules of the
experimental  data on $\tau$-lepton decays and charmonium spectrum
in the papers \cite{Ioffe} were obtained the following values
of gluon condensate:
$\lan\frac{\alpha_s}{\pi}G^2\ran=(0.006\pm 0.012)$ GeV$^4$ and
$\lan\frac{\alpha_s}{\pi}G^2\ran=(0.009\pm 0.007)$ GeV$^4$
correspondingly.

One can assume that the sum rules operate with the some effective
value of the condensate, which equals to the correlator
  $\lan\frac{\alpha_s}{\pi}F_{\mu\nu}^a(x_{\mr{ef}})
F_{\mu\nu}^a(0)\ran$ at some effective separation $x_{\mr{ef}}\ne 0$.
Taking
$\lan\frac{\alpha_s}{\pi}G^2\ran=0.007$~GeV$^4$ and $T_g\approx
0.13$ fm, one obtains  $x_{\mr{ef}}=T_g\ln\left(
\lan\frac{\alpha_s}{\pi}F^2\ran/
\lan\frac{\alpha_s}{\pi}G^2\ran\right) \approx 0.5$ fm, i.e.
characteristic radius of confinement.
Note that when assuming that bilocal correlators do not depend on
point separation and $D(x)$=const, and integrating in (\ref{B17})
up to the characteristic hadronic size $r_h=1$ fm, one can estimate
the gluon condensate from  (\ref{B17}), (\ref{B18}) as
 $\lan\frac{\alpha_s}{\pi}F^2\ran\approx \sigma/r_c^2\approx
0.007$ GeV$^4$, which is also consistent with the value obtained in
 \cite{Ioffe}.

The conclusion is that the sum rules results can not be used neither
as a confirmation nor as a disclaimer of the MFC parameterization
 (\ref{B14}).

\section{The baryon potential}

According to
(\ref{B5}), (\ref{B12}), (\ref{B14}) the baryon potential reads as
$$
V_B=\lim_{T\to \infty}   \frac12
\sum_{a=1,2,3}\int_{S_a}\int_{S_a}
d\sigma_{\mu_1\nu_1}^{(a)}(x)d\sigma_{\mu_2\nu_2}^{(a)}(x')
{\cal D}_{\mu_1\nu_1,\mu_2\nu_2}(x-x')-
$$
\be
 -\frac12
\sum_{a<b}\int_{S_a}\int_{S_b}
d\sigma_{\mu_1\nu_1}^{(a)}(x)d\sigma_{\mu_2\nu_2}^{(b)}(x')
{\cal D}_{\mu_1\nu_1,\mu_2\nu_2}(x-x'),
\label{19}
\ee
where $d\sigma^{(a)}$ denotes an integration over the surface $S_a$.
Note that the nondiagonal part of the potential
 (i.e. second term in (\ref{19}) at $a\ne b$) differs in the factor
 $-1/2=-1/(N_c-1)$ from the corresponding quantity in
\cite{fields,strings}, where the error was made by the author (in
eqs.(24) from \cite{fields} and (46) from \cite{strings}).

Since the surfaces of the Wilson loop are oriented along the temporal
axis, only correlators of the color-electric field contribute
to the potential (\ref{19}), and one gets
\be
V_B(R_1,R_2,R_3)=
\left(\sum_{a=b}-\sum_{a<b}\right)n_i^{(a)}n_j^{(b)}\int_0^{R_a}\int_0
^{R_b } d l\,d l' \int_0^\infty  d t\, {\cal D}_{i4,j4}(z_{ab}),
\label{21}
\ee
where $\ven^{(a)}$ is a vector of unit length directed along the line
connecting string junction with the corresponding quark;
 $z_{ab}=(l\,\ven^{(a)}-l'\ven^{(b)},t)$,
$R_a$ is a distance from the string junction to the corresponding
quark, and
\be {\cal D}_{i4,j4}(z)
=\delta_{ij}D(z)+\frac{\partial}{\partial z_i}\frac{z_j D_1(z)}2.
\label{22}
\ee
In what follows we neglect the formfactor $D_1$ in (\ref{22}), since
its contribution to the potential is small.

In calculation of the baryon potential one should distinguish
configurations of quarks, in which quarks form 1) triangles with each
of the angles less that $2\pi/3$ and 2) triangles having one angle
greater than $2\pi/3$. In the former case the minimal surface of the
Wilson loop consists of three surfaces intersecting with each other at
angles $2\pi/3$, as is shown in Fig. 1. In the latter one  the string
junction  position  coincide with the  position of quark at large
angle, and the Wilson loop consists of two surfaces. Then one can set
in Eq. (\ref{21}) $R_3=0$.

\begin{figure}[!t]
\epsfxsize=12cm
\vspace{-3mm}
\hspace*{2.35cm}
\epsfbox{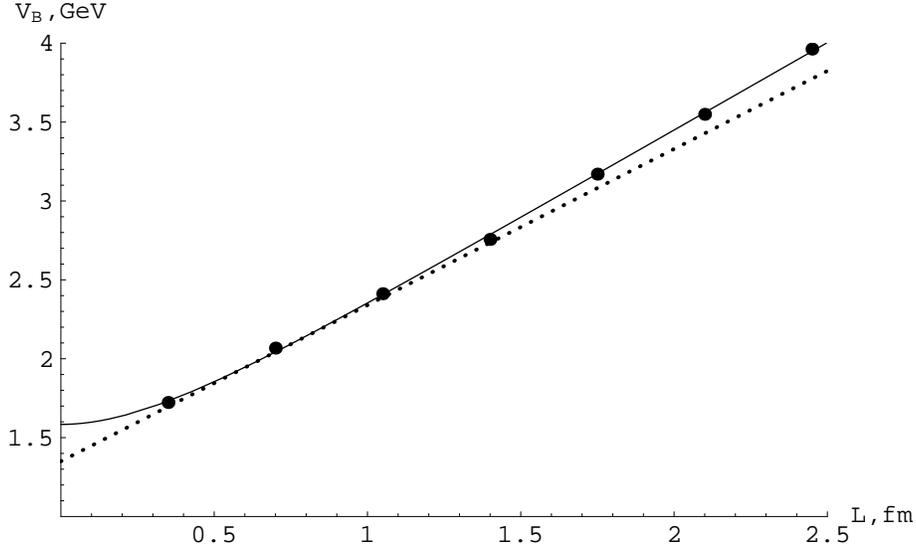}
\caption{The lattice nonperturbative baryon potential
from \cite{Takahashi} (points) for lattice parameter $\beta=5.8$ and  MFC
potential $V^{(B)}$ (solid line) with parameters   $\sigma=0.22$
GeV$^2$ and  $T_g=0.12$ fm vs. the minimal length of the string $L$.
The dotted line is a tangent at $L=0.7$ fm.}
\label{fig2}
\end{figure}

Consider in advance configuration of the equilateral triangle. In this
case $R_1=R_2=R_3\equiv R$, and baryon potential reads as
\be
   V_B(R)=3V_M(R)+V_{\mr{nd}}(R),
\label{24}
\ee
where $V_M$ is diagonal and $V_{\mr{nd}}$ nondiagonal terms.
 From eqs.(\ref{B14}), (\ref{21}),(21) one gets
$$
V_M(R)=2 D(0)\int_0^R d\,z_1(R-z_1)\int_0^\infty d\,t
\exp\lt(-\fr{|z|}{T_g}\rt)=
$$
\be
=\frac{2\sigma}{\pi}\left\{
R\int_0^{R/T_g}d\,x\, x K_1(x)-T_g
\left(2-\frac{R^2}{T_g^2}K_2\left(\frac R{T_g}\right)\right)\right\},
\label{25}
\ee
where $\sigma=\pi D(0) T_g^2$,
$K_1$ and $K_2$ are McDonald functions.
This potential determines an interaction of static quark and antiquark
at separation $R$. Note that it is  related  with the gluon field
in meson calculated in MFC \cite{strings} using the connected probe
as follows,
\be
\frac{d V_M(R)}{d R}=\frac{2\sigma}{\pi}\int^{R/T_g}_0 dx\, x\, K_1(x)
\equiv E_0(R),
\label{26}
\ee
where $E_0(R)$ is a value of the confining field acting on the quark.

The nondiagonal potential,
\be
V_{\mathrm{nd}}(R)=\frac2{\sqrt{3}}\sigma T_g-
\frac{3\sqrt{3}}{2\pi}\frac{\sigma R^2}{T_g}
\int_{\frac{\pi}6}^{\frac{\pi}3}\frac{d\,\varphi}{\cos \varphi}
 K_2\left(\frac{\sqrt{3}R}{2T_g\cos \varphi}\right),
\label{27}
\ee
is positive, rises from zero to the value $2/\sqrt{3}\,\sigma T_g$,
and saturates at $R\ga 0.6$ fm.

The behavior of the potential (\ref{24}) versus the minimal length of
the baryon string $L=R_1+R_2+R_3=3R$ is shown in Fig. 2 in comparison
with the lattice data from \cite{Takahashi}. One can see that the
potential completely describes the lattice results. At large distances
$L\ga 1.5$ fm the potential rises linearly with the slope $\sigma$.
At characteristic hadronic sizes the slope of the potential falls,
which is shown in Fig. 2 by the tangent at point $L=0.7$ fm with the
slope $\sim 0.9 \sigma$. Note that all lattice points at $L<1.5$ fm
are cosistent with the tangent. This effect is also confirmed
by the phenomenology of baryon spectrum \cite{slope}.
\begin{figure}[!t]
\epsfxsize=12cm
\vspace{-3mm}
\hspace*{2.35cm}
\epsfbox{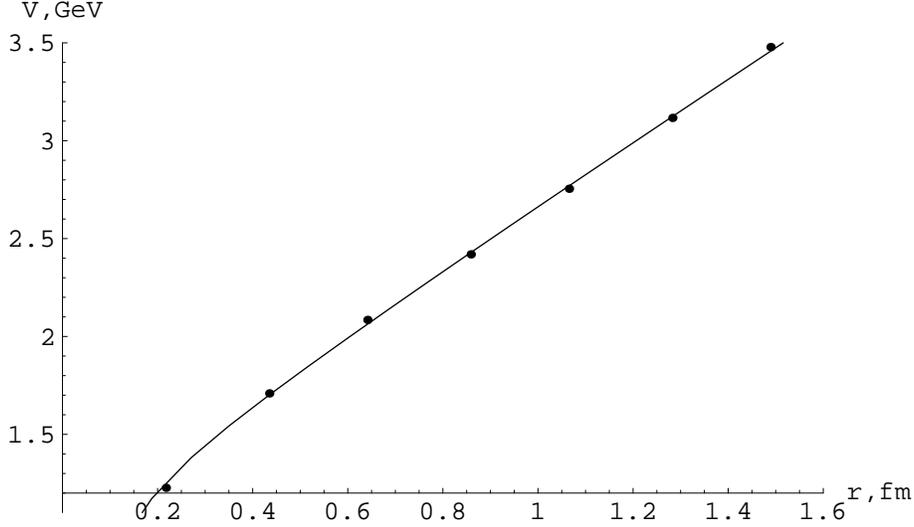}
\caption{The lattice baryon potential in the equilateral triangle with
quark separations $r$ from \cite{deForcrand} (points) at $\beta=5.8$
and the MFC potential $V^{(B)}+V^{\mathrm{pert}}_{\mathrm{(fund)}}$
(solid line) at  $\alpha_s=0.18$, $\sigma=0.18$ GeV$^2$, and
$T_g=0.12$ fm.}
\label{fig3}
\end{figure}

To perform the analysis of the lattice results
from \cite{deForcrand}, one should add to the potential (\ref{24})
the perturbative color-coulomb one,
\be
V_{\mr{pert}}=-\frac32\,\frac{C_F\alpha_s}{r},
\label{28}
\ee
where $C_F=4/3$ is the fundamental Casimir operator. The results
are shown in Fig. 3. One can see the full agreement of our potential
with this independent set of lattice data as well.

In the case of the triangle with different separations $R_a$ of
quarks from the string junction, having angles no larger than
$2\pi/3$ \be
 V_B(R_1,R_2,R_3)=\sum_{a=1}^3V_M(R_a)+
\sum_{a<b}V_\mathrm{nd}(R_a,R_b),
\label{29}
\ee
where
$$
V_{\mr{nd}}(R_a,R_b)=\fr{2\sigma T_g}{3\sqrt{3}}-
\fr{\sqrt{3}\sigma}{4\pi T_g}\lt\{R_a^2\int_0^{\tilde\varphi_{ab}}
\fr{d\,\varphi}{\cos^2\lt(\varphi+\fr\pi 6\rt)}
K_2\lt(\fr{\sqrt{3}R_a}{2T_g\cos\lt(\varphi+\fr\pi 6\rt)}\rt)+\rt.
$$
\be
\lt.+R_b^2\int_{\tilde\varphi_{ab}}^{\fr\pi 3}
\fr{d\,\varphi}{\sin^2\varphi}
K_2\lt(\fr{\sqrt{3}R_b}{2T_g\sin\varphi}\rt)\rt\}.
\label{30}
\ee

If one of the angles of the triangle is larger than $2\pi/3$, the
position of string junction coincides with the position of quark at
large vertex. Let us denote $\alpha$ an
angle supplementary to the large one; $r_1,~r_2$ the distances
from the string junction to the adjacent
 quarks. Then we get
\be
 V_B(r_1,r_2,\alpha)=V_M(r_1)+V_M(r_2)+
V_\mr{nd}(r_1,r_2,\alpha),
\label{31}
\ee
where
$$
V_{\mr{nd}}(r_1,r_2,\alpha)=
\fr{2\alpha\,\mr{ctg}\,\alpha}{\pi}\,\sigma T_g-
\fr{\sin2\alpha\,\sigma}{2\pi T_g}\lt\{r_1^2\int_0^{\tilde\varphi}
\fr{d\,\varphi}{\sin^2(\alpha-\varphi)}
K_2\lt(\fr{r_1\sin\alpha}{T_g\sin(\alpha-\varphi)}\rt)+\rt.
$$
\be
\lt.+r_2^2\int_{\tilde\varphi}^\alpha
\fr{d\,\varphi}{\sin^2\varphi}
K_2\lt(\fr{r_2\sin\alpha}{T_g\sin\varphi}\rt)\rt\}.
\label{32}
\ee
Note that
$V_{\mr{nd}}(r_1,r_2,\alpha)=V_{\mr{nd}}(r_2,r_1,\alpha)$; \quad
$V_{\mr{nd}}(r_1,r_2,\pi/3)=V_{\mr{nd}}(r_1,r_2)$, \quad
$V_{\mr{nd}}(R,R)=1/3!\,V_{\mr{nd}}(R)$.

For the analysis of the dependence of potential on the location of
quarks, let us consider its behavior in isoceles triangles with the
vertex $0\leq\gamma\leq$ $\leq 2\pi/3$ and fixed length of the string
 $L=R_1+R_2+R_3$ versus the value of $\gamma$. At distances
 $R_a\ga 0.5$ fm one can use asymptotic formulas following from
(\ref{25}), (\ref{30}) and (\ref{32}),
$$
V_M(R)\approx \sigma R-\frac4{\pi}\,\sigma T_g,
$$
\be
V_{\mr{nd}}(R_a,R_b)\approx\fr{2\sigma T_g}{3\sqrt{3}},\qquad
V_{\mr{nd}}(r_1,r_2,\alpha)\approx
\fr{2\alpha\,\mr{ctg}\,\alpha}{\pi}\,\sigma T_g.
\label{33}
\ee
One can conclude that at $\gamma=0$, when positions of two quarks
coincide and these quarks are in the antitriplet of color $SU(3)$,
the baryon potential in isoceles triangle reduces to the meson one,
\be
V^L(\gamma=0)\approx \sigma L-\frac4{\pi}\,\sigma T_g.
\label{34}
\ee
If $0<\gamma<2\pi/3$, then the string consists of three lengths
and
\be
V^L(0<\gamma<2\pi/3)\approx \sigma L+\left(-\frac{12}{\pi}
+\fr{2}{\sqrt{3}}\right)\sigma T_g.
\label{35}
\ee
If the string consists of two lengths, then
\be
V^L(\gamma\geq 2\pi/3)\approx \sigma L+\left(-\frac8{\pi}
+\fr2{\pi}(\pi-\gamma)\,\mr{ctg}\,(\pi-\gamma)\right)\sigma T_g.
\label{36}
\ee
According to eqs. (\ref{34}) -- (\ref{36}), when $\gamma$ icreasing,
the potential rapidly falls by the value
\be
\Delta V_1=\left(\fr8{\pi}-\fr2{\sqrt{3}}\right)\sigma T_g
\approx 150 \mbox{ MeV},
\label{37}
\ee
almost does not change in the range $0\la\gamma\la 2\pi/3$, then
rapidly increases at $\gamma\approx 2\pi/3$ by the value
\be
\Delta V_2=\left(\fr4{\pi}-\fr{4}{3\sqrt{3}}\right)\sigma T_g
\approx 55 \mbox{ MeV},
\label{38}
\ee
and slowly  increases  in the range $2\pi/3\leq\gamma\leq\pi$
by the value
\be
\Delta V_3\equiv V^L(\pi)-V^L\left(\fr{2\pi}3\right)=
\left(\fr2{\pi}-\fr{2}{3\sqrt{3}}\right)\sigma T_g
\approx 30 \mbox{ MeV},
\label{39}
\ee
where numerical values are given for $\sigma=0.18$ GeV$^2$,
$T_g=0.12$~fm.
\begin{figure}[!t]
\epsfxsize=12cm
\vspace{-3mm}
\hspace*{2.35cm}
\epsfbox{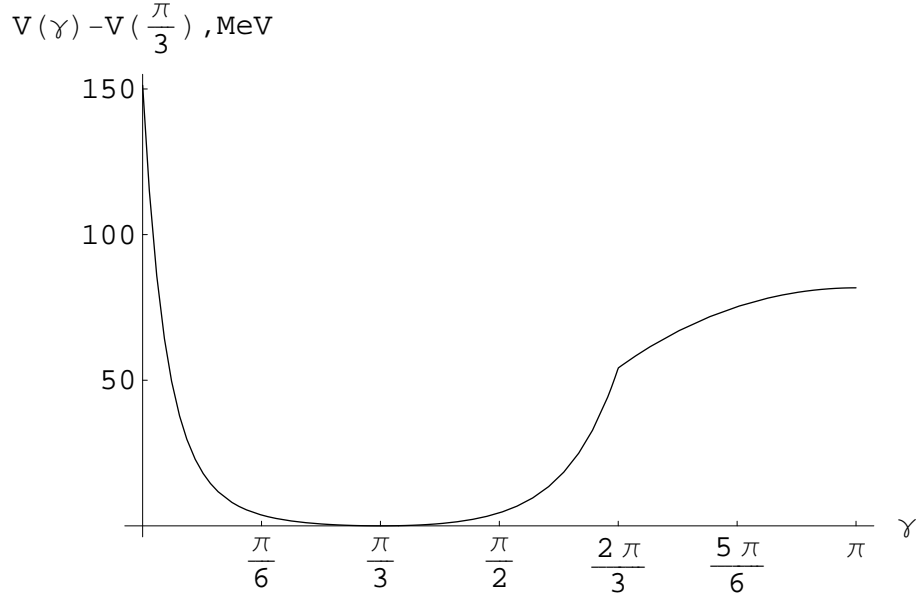}
\caption{The baryon potential in isoceles triangle with the length of the
string $L=1.8$ fm versus the vertex $\gamma$ for $\sigma=0.18$
GeV$^2$, $T_g=0.12$ fm.}
\label{fig4}
\end{figure}
The dependence of the baryon potential in isoceles triangle on the
angle $\gamma$ at $L=1.8$ fm is shown in Fig. 4. Note that character
jumps and drops of the potential are proportional to the quantity
 $\sigma T_g$, i.e. determined by confinement as well as by
correlations of the stochastic nonperturbative fields, and it would be
interesting to obtain them independently in lattice QCD. Baryon
potential in various configurations of quarks was considered on the
lattice in \cite{Takahashi}. However, configurations with the angles
in the range $\pi/20\la\alpha<\pi/2$ considered in this work did not
allow to establish this dependence, because  in the range of the small
angles the size of lattice spacing becomes comparable with the quark
separation and accurate calculations of the potential become
difficult. The further lattice investigations of configurations
with large angle are called for.

It is also interesting to consider the energy
density of the string near the string junction, or the quantity
\be
\tilde V^L(\gamma)=\frac12\left( \sum_{a=1}^3V_M(R_a)-\sigma L\right)+
\sum_{a<b}V_\mathrm{nd}(R_a,R_b).
\label{40}
\ee
If the vertex in the isoceles triangle with the string length fixed
is in the region $2\pi/3\la\gamma\la\pi$, the potential $\tilde
V^L(\gamma)$ has the same behavior as the $V^L(\gamma)$, since in this
region the dependence of the latter on the value of vertex is
determined by the only  nondiagonal term. This means that while
the vertex is increasing in this range, the energy density is rising.
The drop at $\gamma=0$ for  $\tilde V$ is small,
\be
\Delta \tilde V_1=\left(\fr4{\pi}-\fr{2}{\sqrt{3}}\right)\sigma T_g
=0.12 \sigma T_g
\approx 13 \mbox{ MeV}.
\label{41}
\ee
At $\gamma=2\pi/3$ the potential $\tilde V^L(\gamma)$ drops but not
jumps as  $V^L(\gamma)$ does,
\be
\Delta \tilde V_2=\left(\fr2{\pi}-\fr{4}{3\sqrt{3}}\right)\sigma T_g
=-0.13 \sigma T_g
\approx -15 \mbox{ MeV},
\label{42}
\ee
i.e. the energy density decreases. One can conclude that the energy
density near the string junction depends on the location of the quarks
only slightly. This is in contradiction with the field distributions
obtained in \cite{fields,strings} using the connected probe, where
author has made an error in the basic formulas (11)-(18) from
\cite{fields} and (27) from \cite{strings}. It is not hard to show
that the cited formulas and corresponding field distributions concern
to the two or three  static meson Wilson loops in the special case
when the positions of antiquarks coincide, but not to baryons.
The problem of the field distributions in baryon with the connected
probe has some ambiguities and will be considered in subsequent
publications.

\section{Conclusions}

In the present work the nonperturbative static potential in baryon
was calculated analytically in the framework of MFC
for arbitrary locations of quarks.
Two important effects related with the correlations of the
nonperturbative gluon fields are studied: a decrease of the slope of
the potential at hadronic lengths, and its dependence on the locations
of quarks when length of the string fixed. The latter effect is found
for the first time and shown to be
proportional to the combinations of parameters $\sigma T_g$.
It would be interesting to verify it independently on the lattice
and extract an accurate value of the correlation length of gluonic
fields.

The errors in preceding works
\cite{fields,strings} were corrected for baryon potential and stated
for the field distributions in baryon with the connected probe.

 The dependence of the potential on the length of the string was
shown in the paper to be
 in complete agreement with the precise numerical lattice
calculations, which allows to perform the accurate analytic studies
of the baryon spectra, first of all of the spectrum of nucleons, and
has important meaning for various applications in nuclear physics.

The author is grateful to Yu.A.Simonov for numerous useful
discussions. I also thank T.Takahashi, Ph.de Forcrand and
V.I.Shevchenko for the correspondence. Partial support by
 RFBR grants 00-02-17836,  00-15-96786, and INTAS 00-00110, 00-00366
is acknowledged.


\begin{thebibliography}{99}
\looseness=-1

\bibitem{correlators}
 A. Di Giacomo, H.G. Dosch, V.I. Shevchenko and Yu.A. Simonov,
hep-ph/0007223.

\bibitem{MFC}
  H.G. Dosch, Phys.Lett. {\bf B190}, 177  (1987);\\
  H.G. Dosch and Yu.A. Simonov, Phys.Lett. {\bf B205}, 399 (1988);\\
   Yu.A. Simonov, Nucl.Phys. {\bf B307}, 512  (1988).


\bibitem{Regge}
A.Yu. Dubin, A.B. Kaidalov, and Yu.A. Simonov,
Phys.Lett. {\bf B323}, 41 (1994); Phys.Lett. {\bf B343}, 310
(1995);\\ Yu.S. Kalashnikova, A.V. Nefediev, and Yu.A. Simonov,
Phys.Rev. {\bf D64}, 014037 (2001).

\bibitem{glump} Yu.A. Simonov, Nucl.Phys. {\bf B592}, 350 (2000).

\bibitem{Takahashi} T.T. Takahashi {\it et al.}, Phys.Rev. {\bf D65},
114509 (2002).

\bibitem{deForcrand} C. Alexandrou, Ph. de Forcrand, and  O. Jahn,
 hep-lat/0209062, talk presented at Lattice'2002.


\bibitem{Gauss}
Yu.A.Simonov, ðÉÓØÍÁ × öüôæ {\bf 71}, 187 (2000), hep-ph/0001244;\\
V.I.Shevchenko and Yu.A.Simonov, Phys.Rev.Lett. {\bf 85}, 1811 (2000).

\bibitem{Casimir}
G.S.Bali, Nucl.Phys. B (Proc.Suppl.) {\bf 83}, 422 (2000).

\bibitem{cluster}
N.G. Van Kampen, Stochastic Processes in Physics and Chemistry,
North-Holland Physics Publishing, 1984.


\bibitem{latDD1}
   M. Campostrini, A. Di Giacomo and G. Mussardo, Z.Phys. {\bf C25},
173    (1984);\\
A. Di Giacomo and H. Panagopoulos, Phys.Lett. {\bf B285 }, 133
(1992);\\ A. Di Giacomo, E. Meggiolaro and H. Panagopoulos, Nucl.Phys.
 {\bf B483 }, 371 (1997).

\bibitem{BBV}
G.S. Bali, N. Brambilla, and A. Vairo,
Phys.Lett. {\bf B421}, 265 (1998).

\bibitem{glump}
Yu.A. Simonov, Nucl.Phys. {\bf B592}, 350 (2001).

\bibitem{zero}
Yu.A. Simonov, Sov.J.Nucl.Phys. {\bf 50}, 134 (1989).

\bibitem{baryon}
Yu.A. Simonov, hep-ph/0205334.

\bibitem{sumrules}F.J. Yndurain, The Theory of Quark and Gluon
Interactions, Springer-Verlag, 1999, p. 208.

\bibitem{Ioffe} B.V. Geshkenbein, B.L. Ioffe, K.N. Zyablyuk, Phys.Rev.
{\bf D64,} 093009 (2001);\\B.L. Ioffe, K.N. Zyablyuk,
hep-ph/0207183;\\ B.L. Ioffe, hep-ph/0209313.


\bibitem{fields}
D.S. Kuzmenko and Yu.A. Simonov, Phys.Lett. {\bf B494}, 81 (2000).

\bibitem{strings}
D.S. Kuzmenko and Yu.A. Simonov, Yad.Fiz. {\bf 64}, 110 (2001),
hep-ph/0010114.

\bibitem{slope}
S. Capstick, N. Isgur, Phys.Rev. D {\bf 34,} 2809 (1986).

\end{thebibliography}
\end{document}